%% file: main.tex
\documentclass[12pt,a4paper]{article}

\usepackage[utf8]{inputenc}
\usepackage[english]{babel}
\usepackage{fullpage}
\usepackage{pdflscape}
\usepackage{amsmath}
\usepackage{amssymb}
\usepackage{amsthm}
\usepackage{fixmath}

\newcommand{\ar}{\text{Ar}}
\newcommand{\fv}{\text{FV}}
\newcommand{\amb}{\text{Amb}}

\title{Macro Lambda Calculus}
\author{Anton Salikhmetov}

\begin{document}
\maketitle

\begin{abstract}
The goal of our Macro Lambda Calculus project (MLC) is to encode $\lambda$-terms into interaction nets.
Its software implementation accepts input in the notation similar to $\lambda$-calculus allowing macro definitions.
Output is similar to interaction calculus and is suitable for our Interaction Nets Compiler program (INC).
In this paper, we describe the interaction system for call-by-need evaluation and the mechanism of encoding $\lambda$-terms into this system which MLC is based on.
\end{abstract}

\input workflow
\input amb
\input encoding

\input biblio
\newpage
\appendix
\input rules
\end{document}

%% file: workflow.tex
\section{Workflow}

MLC exists in the context of a set of tools needed to evaluate $\lambda$-terms.
A $\lambda$-term to be reduced to its normal form (if such exists) is written down using textual representation to be further translated into an interaction net by the MLC compiler \texttt{mlc}.
In the spirit of UNIX command pipelines, the \texttt{mlc} output can be piped into the interaction net compiler \texttt{inc}.
The \texttt{inc} compiler translates the interaction net into a C language program that will perform interaction.
In our case, this program (once compiled into binary) computes and prints out the normal form of the original $\lambda$-term if such exists or enters an infinite loop otherwise.
The output syntax is the same as the one used to encode the original term.

%% file: amb.tex
\section{Non-deterministic extension}

We work in interaction calculus~\cite{calculus} extended by a special non-deterministic agent $\amb$~\cite{amb}. We represent this agent in a more conservative fashion than it was suggested in the original paper. Specifically, we prepend the list of auxiliary ports of $\amb$ with its extra principal port and introduce the following conversion:
$$
\langle\vec t\ |\ t = \amb(u, v, w), \Delta\rangle 
=
\langle\vec t\ |\ u = \amb(t, v, w), \Delta\rangle.
$$
We assume that any interaction system's signature $\Sigma$ is implicitly extended by $\amb$ with ${\ar(\amb) = 3}$, while its set of rules is implicitly extended with
$$
\alpha[\vec x] \bowtie \amb[y, \alpha(\vec x), y].
$$

%% file: encoding.tex
\section{Encoding $\lambda$-terms}

We denote the set of $\lambda$-terms~\cite{lambda} as $\Lambda$, and $C[\phantom M]$ means a context, i.~e.~a $\lambda$-term with one hole, while $C[M]$ is the result of placing $M$ in the hole of the context $C[\phantom M]$.

Our interaction system has signature
$$
\Sigma = \{\epsilon, s, @, \lambda, c, \delta\}
\cup
\{a_M\ |\ M \in \Lambda\}
\cup
\{r_{C[\phantom M]}\ |\ \text{$C[\phantom M]$ is a context}\},
$$
with all the agents being binary, except ${\ar(\epsilon) = \ar(a_M) = 0}$ and ${\ar(r_{C[\phantom M]}) = 1}$.
For the list of interaction rules, please refer to Appendix~\ref{rules} which also includes a tabular index. 

While encoding $\lambda$-terms into our interaction system, we will distinguish their free variables from their bound variables.
So, let us mark all free variables in a $\lambda$-term $M$ using the following operation: ${M^\bullet \equiv M[\vec x := \vec x^\bullet]}$, where ${(\vec x) = \fv(M)}$.
Then, any $\lambda$-term $M$ can be mapped to configuration ${\langle x\ |\ r_{[\phantom M]}(x) = y, \Gamma(M^\bullet, y)\rangle}$ as follows:
\begin{align*}
\Gamma(x^\bullet, y) &= \{a_x = y\}; \\
\Gamma(x, y) &= \{x = y\}; \\
\Gamma(\lambda x.M, y) &= \{y = \lambda(\epsilon, z)\} \cup \Gamma(M, z), \quad \text{when}\ x \not\in \fv(M); \\
\Gamma(\lambda x.M, y) &= \{y = \lambda(x, z)\} \cup \Gamma(M, z), \quad \text{when}\ x \in \fv(M); \\
\Gamma(M\ N, y) &= \{y = @(x, z)\} \cup \Gamma(M[\vec t := \vec t'], x) \cup \Gamma(N[\vec t := \vec t''], z) \cup \Psi(\vec t), \quad \text{where} \\
\Psi(\vec t) &= \{t'_i = \amb(t''_i, s(t_i, u_i), u_i)\ |\ t_i \in (\vec t)\}, \quad \text{and} \\
(\vec t) &= \fv(M) \cap \fv(N).
\end{align*}
We claim that ${\langle x\ |\ r_{[\phantom M]}(x) = y, \Gamma(M^\bullet, y)\rangle \downarrow \langle a_N\ |\ \varnothing\rangle}$ iff ${M \twoheadrightarrow N}$ and $N$ is normal form.

Please note that our interaction system has read-back mechanism embedded.
Indeed, configuration that encodes a $\lambda$-term $M$ will be reduced to normal form (if any) with only one agent $a_N$ in its interface, $N$ representing normal form of the encoded $\lambda$-term $M$.

%% file: rules.tex
\section{Interaction rules}
\label{rules}

\begin{table}
\centering
\begin{tabular}{c|cccccccc|c}
&
$\epsilon$ &
$a$ &
$s$ &
$@$ &
$\lambda$ &
$c$ &
$\delta$ &
$r$ &
\\

\hline

$\epsilon$ &
(\ref{epsilon-epsilon}) &
(\ref{epsilon-a}) &
(\ref{epsilon-s}) &
(\ref{epsilon-@}) &
(\ref{epsilon-lambda}) &
(\ref{epsilon-c}) &
(\ref{epsilon-delta}) &
--- &
$\epsilon$ \\

$a$ &
(\ref{epsilon-a})&
--- &
--- &
--- &
(\ref{a-lambda}) &
(\ref{a-c}) &
(\ref{a-delta}) &
(\ref{a-r}) &
$a$ \\

$s$ &
(\ref{epsilon-s}) &
--- &
--- &
--- &
(\ref{s-lambda}) &
(\ref{s-c}) &
(\ref{s-delta}) &
(\ref{s-r}) &
$s$ \\

$@$ &
(\ref{epsilon-@})&
--- &
--- &
--- &
(\ref{@-lambda}) &
(\ref{@-c}) &
(\ref{@-delta}) &
(\ref{@-r}) &
$@$ \\

$\lambda$ &
(\ref{epsilon-lambda}) &
(\ref{a-lambda}) &
(\ref{s-lambda}) &
(\ref{@-lambda}) &
(\ref{lambda-lambda}) &
(\ref{lambda-c}) &
(\ref{lambda-delta}) &
(\ref{lambda-r}) &
$\lambda$ \\

$c$ &
(\ref{epsilon-c}) &
(\ref{a-c}) &
(\ref{s-c}) &
(\ref{@-c}) &
(\ref{lambda-c}) &
--- &
(\ref{delta-c}) &
--- &
$c$ \\

$\delta$ &
(\ref{epsilon-delta}) &
(\ref{a-delta}) &
(\ref{s-delta}) &
(\ref{@-delta}) &
(\ref{lambda-delta}) &
(\ref{delta-c}) &
(\ref{delta-delta}) &
--- &
$\delta$ \\

$r$ &
--- &
(\ref{a-r}) &
(\ref{s-r}) &
(\ref{@-r}) &
(\ref{lambda-r}) &
--- &
--- &
--- &
$r$ \\

\hline

&
$\epsilon$ &
$a$ &
$s$ &
$@$ &
$\lambda$ &
$c$ &
$\delta$ &
$r$ &
\end{tabular}
\end{table}
\begin{align}
\epsilon &\bowtie \epsilon; \label{epsilon-epsilon} \\
\epsilon &\bowtie a_M; \label{epsilon-a} \\
\epsilon &\bowtie s[x, x]; \label{epsilon-s} \\
\epsilon &\bowtie @[\epsilon, \epsilon]; \label{epsilon-@} \\
\epsilon &\bowtie \lambda[\epsilon, \epsilon]; \label{epsilon-lambda} \\
\epsilon &\bowtie c[\epsilon, \epsilon]; \label{epsilon-c} \\
\epsilon &\bowtie \delta[\epsilon, \epsilon]; \label{epsilon-delta} \\
a_M &\bowtie \lambda[r_{M\ [\phantom M]}(x), x]; \label{a-lambda} \\
a_M &\bowtie c[a_M, a_M]; \label{a-c} \\
a_M &\bowtie \delta[a_M, a_M]; \label{a-delta} \\
a_M &\bowtie r_{C[\phantom M]}[a_{C[M]}]; \label{a-r} \\
s[c(z, \lambda(x, y)), z] &\bowtie \lambda[x, y]; \label{s-lambda} \\
s[c(z, c(x, y)), z] &\bowtie c[x, y]; \label{s-c} \\
s[c(x, \delta(y, z)), x]  &\bowtie \delta[y, z]; \label{s-delta} \\
s[c(x, r_{C[\phantom M]}(y)), x] &\bowtie r_{C[\phantom M]}[y]; \label{s-r} \\
@[\lambda(z, \lambda(x, y)), z] &\bowtie \lambda[x, y]; \label{@-lambda} \\
@[\lambda(z, c(x, y)), z] &\bowtie c[x, y]; \label{@-c} \\
@[\delta(x, y), \delta(v, w)] &\bowtie \delta[@(x, v), @(y, w)]; \label{@-delta} \\
@[\lambda(x, r_{C[\phantom M]}(y)), x] &\bowtie r_{C[\phantom M]}[y]; \label{@-r} \\
\lambda[x, y] &\bowtie \lambda[x, y]; \label{lambda-lambda} \\
\lambda[\delta(x, y), \delta(v, w)] &\bowtie c[\lambda(x, v), \lambda(y, w)]; \label{lambda-c} \\
\lambda[\delta(x, y), \delta(v, w)] &\bowtie \delta[\lambda(x, v), \lambda(y, w)]; \label{lambda-delta} \\
\lambda[a_y, r_{C[\lambda y.[\phantom M]]}(x)] &\bowtie r_{C[\phantom M]}[x], \quad \text{where ${y \in \Lambda}$ is a new variable}; \label{lambda-r} \\
\delta[x, y] &\bowtie c[\delta(\amb(v, s(x, z), z), \amb(w, s(y, u), u)), \delta(v, w)]; \label{delta-c} \\
\delta[x, y] &\bowtie \delta[x, y]. \label{delta-delta}
\end{align}